\documentclass[preprints,article,accept,moreauthors,pdftex]{mdpi}

\usepackage{graphicx,epstopdf}

\firstpage{1} 
\pubvolume{1}
\issuenum{1}
\articlenumber{0}
\pubyear{2023}
\copyrightyear{2023}
\datereceived{6 December 2022} 
\dateaccepted{31 January 2023} 
\datepublished{} 
\hreflink{https://doi.org/} 
\pdfoutput=1

\Title{Influence of the magnetic field topology in the evolution of small-scale two-fluid jets in the solar atmosphere}

\TitleCitation{Influence of the magnetic field topology in the evolution of small-scale two-fluid jets in the solar atmosphere}

\Author{E. E. Díaz-Figueroa $^{1}$, G. Ares de Parga $^{1}$ and J. J. González-Avilés $^{2}$*\orcidC{}}

\AuthorNames{E. E. Díaz-Figueroa,  G. Ares de Parga and  J. J. González-Avilés}

\AuthorCitation{Díaz-Figueroa, E. E.; Ares de Parga, G.;  González-Avilés, J. J.}

\address{%
$^{1}$ \quad Departamento de Física, Escuela Superior de Física y Matemáticas, Instituto Politécnico Nacional, U. P. Adolfo López Mateos, Zacatenco, C.P. 07738, CDMX, México.; ediazf1001@alumno.ipn.mx, gadpau@hotmail.com\\
$^{2}$ \quad Investigadores por México-CONACYT, SCIESMEX-LANCE, Instituto de Geofísica, Unidad Michoacán, Universidad Nacional Autónoma de México, C.P. 58190 Morelia,
Michoacán, México.; jjgonzalez@igeofisica.unam.mx}

\corres{Correspondence: jjgonzalez@igeofisica.unam.mx}

\abstract{We perform a series of numerical simulations to recreate small-scale two-fluid jets using the JOANNA code, considering the magnetohydrodynamics of two fluids (ions + electrons and neutral particles). We first excite the jets in a uniform magnetic field by using velocity pulse perturbations located at  $y_{0}=$1.3, 1.5, and 1.8 Mm, considering the base of the photosphere at $y=0$ Mm. Then, we repeat the excitation of the jets in a magnetic field that mimics a flux tube. Mainly, the jets excited at the upper chromosphere ($y\sim1.8$ Mm) reach lower heights than those excited at the lower chromosphere ($y\sim1.3$ Mm); this is due to the higher initial vertical location because of the lesser amount of plasma dragging. In both scenarios, the dynamics of the neutral particles and ions show similar behavior; however, we can still identify some differences in the velocity drift, which in our simulations is of the order of $10^{-3}$ km s$^{-1}$ at the tips of the jets once they reached their maximum heights. Also, we estimate the heat generation due to the friction between ions and neutrals ($Q^{in}_{i,n}$), which is of the order of $0.002-0.06$ W m$^{-3}$; however it is small to contribute to the heating of the surroundings of the solar corona. The jets in the two magnetic environments do not show substantial differences other than a slight variation in the maximum heights reached, particularly in the uniform magnetic field scenario. Finally, the maximum heights reached by the three different jets are in the range of some morphological parameters corresponding to macrospicules, Type I spicules, and Type II spicules.}


\begin{document}



\section{Introduction}

Solar jet-type phenomena are ubiquitous in the solar atmosphere. Its importance has generated numerous investigations on its origin and evolution \citep[see, e.g.,][and references therein]{2016SSRv..201....1R, 2021RSPSA.47700217S}. Although there are many open questions about its nature, enormous progress has been made in understanding its dynamics, particularly in 2D, 2.5D, and 3D MHD numerical simulations. These models have swept a broad spectrum of complexity from ideal MHD to resistive MHD and two-fluid MHD \citep[see, e.g.,][]{10.1093/pasj/58.2.423, Pariat_2009, Archontis_et_al_2010, Jiang_2012, Leake_2013, Soler_2013,10.1093/mnras/sty2306}.

Jet-type phenomena have been proposed as promising candidates to generate large amounts of heat at the upper atmospheric layers of the Sun \citep{Beckers_1968}. Their frequency of occurrence is high compared to other solar phenomena, which gives the impression that jets are ongoing events that could represent a continuous source of energy \citep{De_Pontieu_2009}. These collimated plasma jets are called spicules and are cataloged according to their different characteristics \citep{Lippincott_1957, De_Pontieu_2007b, Loboda_2019}. The spicules were first described in \citet{Secchi_1877}, but they owe their name to \citet{Roberts_1945}. Additionally, \citet{Beckers_1968} developed a theoretical and observational analysis to show that the chromosphere is mostly populated by these elongated structures that may supply plasma to the corona. The first spicules observed, also called Type I spicules, can reach heights in the range of 7000-13000 km \citep{Lippincott_1957}, diameters of 500-2000 km \citep{Lynch_1973}, vertical speeds of 25 km s$^{-1}$, lifetimes of 1-10 min, temperatures of 5000-15000 K and densities of 3$\times10^{-13}$ g cm$^{-3}$ that remain quasi-constant with height \citep{Lippincott_1957, Bray_1974}. On the other hand, the Type II spicules, which could be generated due to magnetic reconnection, reach heights of 1000 to 7000 km above the chromosphere, in a range of vertical speeds between 40 km s$^{-1}$ to 300 km s$^{-1}$, with the bulk between 50 and 150 km s$^{-1}$, lifetimes from 10 to 150 s, with characteristic diameters less than 200 km, temperatures of approximately $10^4$ K \citep{De_Pontieu_2007b, Pereira_2012, Isobe_2008, Archontis_et_al_2010, Gonzalez_Aviles_et_al_2017, Gonzalez_Aviles_et_al_2018}. Besides, the macrospicules can reach heights of 7 to 70 Mm and speeds of 10 to 150 km s$^{-1}$, and they can have lifetimes of 3 to 45 min, according to observations and numerical simulations \citep{Murawski_et_al_2011, Loboda_2019, Gonzalez-Aviles_et_al_2021}. Finally, there is also another complex chromospheric ejections, such as the surges, which are seen as darkenings in the blue/red wings of the line with line-of-sight (LOS) apparent velocities of a few to several tens of km s$^{-1}$ on areas with projected lengths of 10–50 Mm \cite{Nobrega-Siverio_2017}. Surges also can be consist of small-scale thread-like structures that appear to related to shocks and Kelvin-Helmholtz instabilities \citep{Nelson&Doyle_2013, Yang_et_al_2014, ZHELYAZKOV20152727}.

The origin of the small-scale plasma jets found in the lower solar atmosphere is still a matter of debate. Therefore, there is still room for new models, such as the two-fluid approximation, which includes the dynamics of neutrals apart from the ions, and therefore is more realistic for studying the generation, evolution, and morphology of small-scale jets from the chromosphere to the solar corona. The two-approximation is essential to modeling of partially ionized plasmas in astrophysical scenarios in general \citep[see, e.g.,][]{Ballester_et_al_2018}. For example, in \cite{Ku_ma_2017}, the authors use numerical simulations using the two-fluid equations in 2D Cartesian geometry to study the formation and evolution of solar spicules. They found that the simulated spicule consists of a dense, cold core dominated by neutrals. More recently, in \cite{Gonzalez-Aviles_et_al_2022}, the authors study the formation and evolution of jets employing localized non-linear Gaussian pulses of ion and neutral pressures initially launched from the magnetic null point of a potential arcade in a partially ionized solar atmosphere. They found that the shock propagates upwards into the solar corona and lifts the cold and dense chromospheric plasma in the form of a collimated jet with an inverted-Y shape. These kinds of inverted-Y jets and their heating may explain the properties of some jets observed in the solar atmosphere. Additionally, there are recent papers related to investigations about the two-fluid effects playing an essential role in the non-linear regime, particularly in the context of wave damping and plasma heating of the solar chromosphere \citep[see, e.g.,][]{Kuzma_et_al_2019, Wojcik_et_al_2020, Murawski_et_al_2020}. 

In this paper, with the use of the JOANNA code \citep{Wojcik_2019}, we solve the two-fluid MHD equations numerically to simulate different chromospheric small-scale two-fluid jets excited at three different vertical locations ($y_{0}=1.3,1.5,1.8$ Mm) to analyze the effect two different magnetic configurations on the evolution and morphology of the jets. We organize the paper as follows. First, section \ref{Model_and_methods} describes the two-fluid equations, the numerical methods, the perturbations, and the magnetic field configurations. Then, in section \ref{Results}, we present the most significant results of the numerical simulations for the two different magnetic field configurations. Next, in section \ref{discussion}, we discuss the most relevant differences between the two simulation cases. Finally, in Section \ref{Conclusions_final_comments}, we draw the results and the conclusions.

\section{Model and methods}
\label{Model_and_methods}

\subsection{The system of two-fluid equations}
\label{System_of_equations}

We consider a stratified solar atmosphere composed of two fluids, i.e., ions+electrons and neutral particles. We write the system of the two-fluid equations as follows \citep{refId0, Oliver_2016, Ku_ma_2017}:

      \begin{eqnarray}
      \label{m_i}
      \frac{\partial \rho_{i}}{\partial t}+\nabla \cdot (\rho_{i} \textbf{V}_{i})
    = 0, \\ 
    \label{m_n}
    \frac{\partial \rho_{n}}{\partial t}+\nabla \cdot (\rho_{n} \textbf{V}_{n})
    = 0, \\
    \label{s_i}
      \frac{\partial \rho_{i}\textbf{V}_{i}}{\partial t}+\nabla \cdot (\rho_{i}\textbf{V}_{i}\textbf{V}_{i}+p_{ie}\textbf{I}) -\frac{1}{\mu}(\nabla \times  \textbf{B}) \times \textbf{B} + \rho_{i} \textbf{g} = -\textbf{S}_{in}, \\ 
      \label{s_n}
    \frac{\partial \rho_{n}\textbf{V}_{n}}{\partial t}+\nabla \cdot (\rho_{n}\textbf{V}_{n}\textbf{V}_{n}+p_{n}\textbf{I}) + \rho_{n} \textbf{g} = \textbf{S}_{ni}, \\
    \label{e_i}
        \frac{\partial p_{i}}{\partial t}+ \textbf{V}_{i} \cdot \nabla p_{i} + \gamma p_{i}  \nabla  \cdot \textbf{V}_{i}  = (\gamma - 1)Q_{i}^{in}, \\
         \label{e_n}
         \frac{\partial p_{n}}{\partial t}+ \textbf{V}_{n} \cdot \nabla p_{n} + \gamma p_{n}  \nabla  \cdot \textbf{V}_{n}  = (\gamma - 1) Q_{n}^{in}, \\
         \frac{\partial \textbf{B}}{\partial t}=\nabla \times (\textbf{V}_{i} \times \textbf{B}), \\
         \label{B} \nabla \cdot \textbf{B}=0,
      \end{eqnarray}
where $\rho_{i,n}$, $p_{i,n}$, $\mu$, $\textbf{B}$, $\textbf{V}_{i,n}$, $\textbf{S}_{in}$, $Q_{i,n}^{in}$; represents the ion ($i$) and neutral ($n$) densities, gas pressures, the magnetic permeability of the medium, magnetic field, the ion and neutral velocities, the collisional momentum and the heat generation due to collisions between species, respectively. But, the electrical resistivity is not included in the simulations. The collisional momentum, $S_{in}$, between particles (ions and neutrals); specifically, in Eq. (\ref{s_i}), $S_{in}$ is defined as $\textbf{S}_{in}=\nu_{in} \rho_{i} (\textbf{V}_{i}-\textbf{V}_{n})$; where the collision frequency between species is $\nu_{in}=\alpha_{in}/\rho_{n}$; and $\alpha_{in}=\frac{4}{3}\frac{\sigma_{in}}{m_{p}+m_{n}} \sqrt{\frac{8k_{B}}{\pi} \left(\frac{T_{i}}{m_{i}}+\frac{T_{n}}{m_{n}}\right)}\rho_{i}\rho_{n}$, where, $\sigma_{in}$, is the collisional cross-section between ions and neutrals and takes a value of $0.75\times10^{-18}$ m$^{2}$ \citep{Oliver_2016}. Finally, the heat generation due to collisions between species, respectively is defined as
\begin{eqnarray}
     Q_{i}^{in}= \alpha_{in} \left[\frac{1}{2}|\textbf{V}_{i}-\textbf{V}_{n}|^{2}+\frac{3}{2}\frac{k_{B}}{m_{i}}(T_{n}-T_{i}) \right], 
     \label{9} \\
     Q_{n}^{in}= \alpha_{in} \left[\frac{1}{2}|\textbf{V}_{i}-\textbf{V}_{n}|^{2}+\frac{3}{2}\frac{k_{B}}{m_{i}}(T_{i}-T_{n}) \right];
     \label{10}
\end{eqnarray}
being $\gamma=5/3$ the adiabatic index; $\textbf{I}$ is a unity matrix. A gravitational acceleration acting only on the $y$-axis are taken ($\textbf{g}=[0,-g]$) with a value equal to $g=$274 m s$^{-2}$. We define the gas pressures by using the ideal gas laws:

\begin{eqnarray}
     p_{i}= \frac{k_{B}}{m_{i}}\rho_{i} T_{i}, \label{11} \\
     p_{n}= \frac{k_{B}}{m_{n}} \rho_{n} T_{n};
     \label{12}
\end{eqnarray}
where $T_{i,n}$ represents the ion and neutral temperatures, respectively; $m_{i}=m_{H}\mu_{i}$, $m_{n}=m_{H}\mu_{n}$, with $m_{H}$ being the hydrogen mass, which is the main ingredient of the gas, and therefore $m_{n}\simeq m_{i}=m_{H}=m_{p}$ (with $m_{p}$ being the proton mass), and $k_{B}$ is the Boltzmann constant. The mean masses $\mu_{i}\approx 0.58$ and $\mu_{n}\approx1.21$. 

\subsection{Model of the solar atmosphere}
\label{mod_sol_atmos}

\begin{figure}[t!]
   \centerline{\includegraphics[width=0.93\textwidth,clip=]{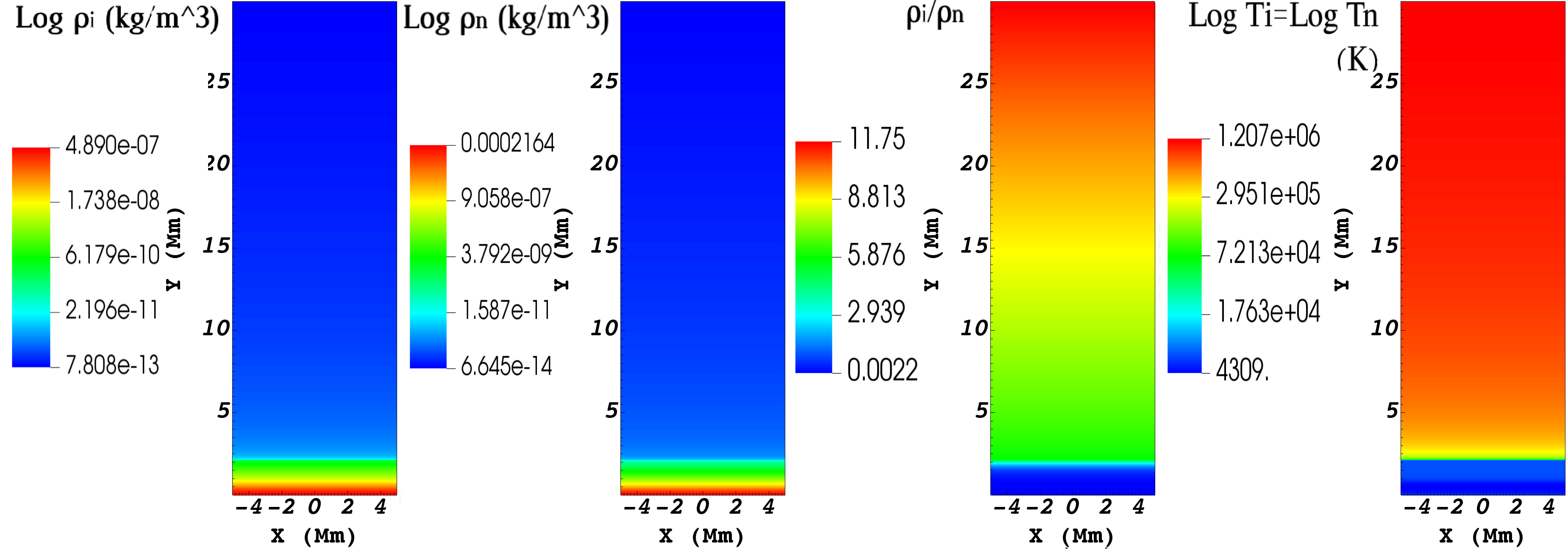}
              }
              \caption{From left to right $\log (\rho_{i})$, $\log (\rho_{n})$, ion-neutral ratio ($\rho_{i}/\rho_{n}$) for the solar atmosphere model at $t=0$ s, and  $\log (T_{i})=\log T_{n})$.}
\label{equilibrium}
\end{figure}

At the initial time of simulations, we assume that the solar atmosphere is in hydrostatic equilibrium, i.e., we set the ion and neutral velocities equal to zero (${\bf V_{i}=V_{n}=0}$). Then, considering the ideal gas laws for ions and neutrals, given by equations (\ref{11})-(\ref{12}), and taking into account the $y$-components of the hydrostatic equation ($-\nabla p_{i,n} + \rho_{i,n}{\bf g} = {\bf 0}$), we arrive to the following expressions for the equilibrium gas pressures:
\begin{eqnarray}
p_{n}(y) &=& p_{0n}\exp{\left(-\int_{y_{0}}^{y}\frac{dy^{\prime}}{\Lambda_{n}(y^{\prime})}\right)}, \label{p_n} \\
p_{i}(y) &=& p_{0i}\exp{\left(-\int_{y_{0}}^{y}\frac{dy^{\prime}}{\Lambda_{i}(y^{\prime})}\right)}. \label{p_i} 
\end{eqnarray}
Here
\begin{eqnarray}
\Lambda_{i}(y) = \frac{k_{B}T_{i}(y)}{m_{H}\mu_{i}g} \quad \mbox{and} \quad \Lambda_{n}(y) = \frac{k_{B}T_{n}(y)}{m_{H}\mu_{n}g} 
\end{eqnarray}
are the pressure scale heights, and $p_{0_{i,n}}$ denote the gas pressures at the reference level $y_{0}=10$ Mm. Specifically, this paper adopts the semi-empirical model of Table 26 of \cite{Avrett_2008} for the temperature field. We consider that temperatures of ions and neutrals are initially equal (at $t=0$ s), i.e., they are in thermal equilibrium, and we set $T_{i}=T_{n}=T$. We also display the equilibrium profiles, including the mass densities for ions and neutrals and the ionization fraction, $\varrho_{i}/\varrho_{n}$ in Fig. \ref{equilibrium}. 

\subsection{Magnetic field configurations}
\label{mag_field_configs}

\begin{figure}[t!]    
 \centering
\includegraphics[width=7.3cm]{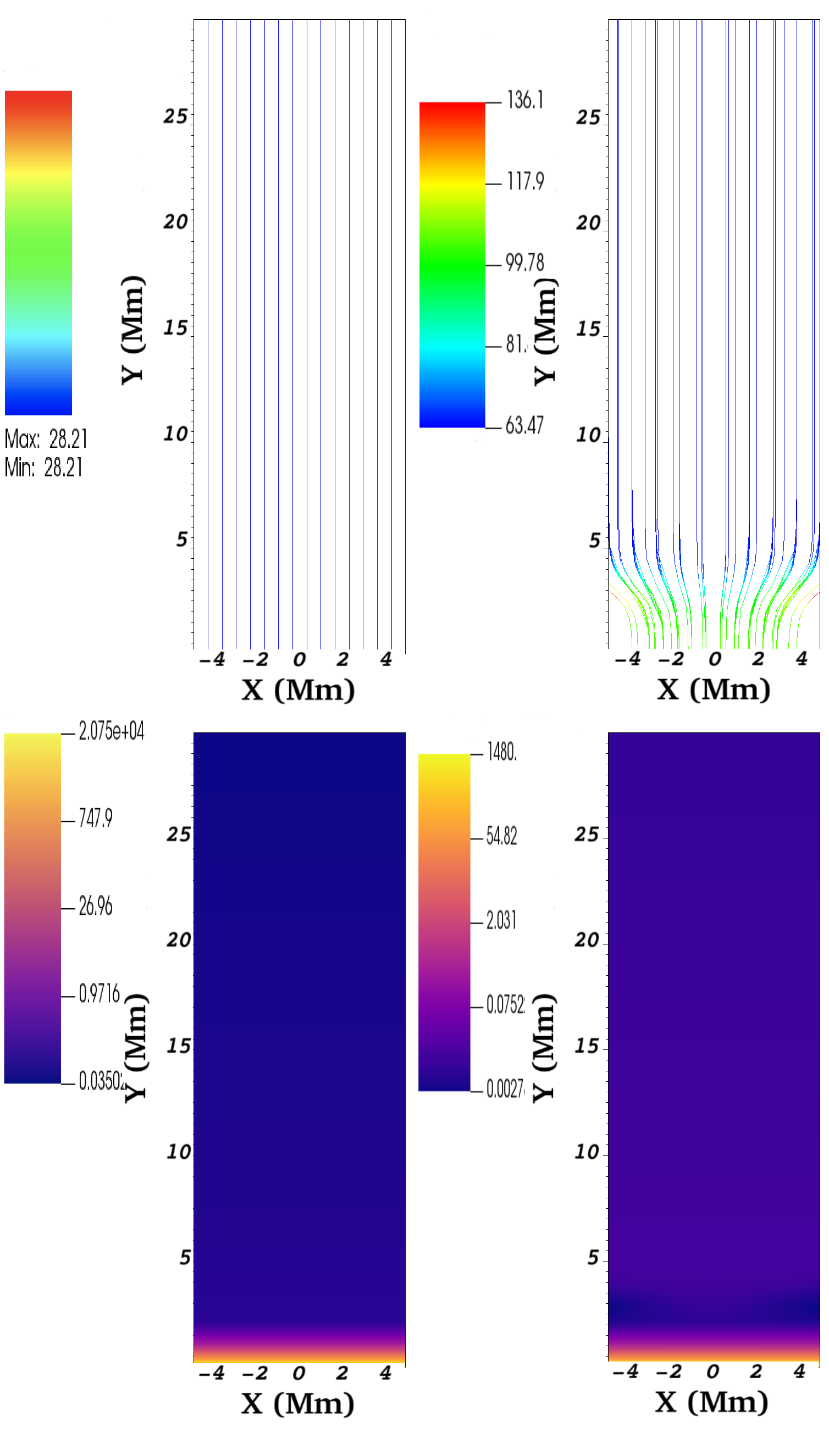}
\caption{\textit{Top}: Magnetic field lines for the vertical straight magnetic field configuration (left) and for the flux tube configuration (right) at the initial time ($t=0$ s). In this figure, the color bar represents the magnitude of the magnetic field $|{\bf B}|$. \textit{Bottom}: The plasma $\beta$ corresponding to the vertical straight magnetic field (left) and the flux tube (right) at the initial time ($t=0$ s).}
   \label{mag_field_plasma_beta}
  \end{figure}

We perform the simulations for two magnetic fields: i) constant vertical field and ii) flux tube. For the constant magnetic field, we use the following:

\begin{equation}
{\bf B} = [0,B_{0}],
\end{equation}
where  $B_{0}=30$ G, displayed in the top-left panel of Fig. \ref{mag_field_plasma_beta}. For the flux tube, we use the following normalized expressions in Cartesian coordinates, which recreate a flux tube in a particular simple geometric way

\begin{equation}
     B_{x}= 0.075\bar{x}B_{0}\textrm{sech}^{2}(\bar{y}-3), \label{Bx}
\end{equation}
 \begin{equation}
     B_{y} = B_{0}(0.3-0.075\textrm{tanh}(\bar{y}-3)). \label{By}
 \end{equation}
where $\Bar{x}=x/L$, $\Bar{y}=y/L$ and $L=10^{6}$ m. Here $B_{0}=100$ G represents the magnitude of the field in the lower part of the flux tube, i.e., in the photosphere. This magnetic field satisfies the divergence-free condition $\nabla \cdot \textbf{B}=0$, and it is analogous to an inverted magnetic bottle that can accelerate to the charged particles at the footpoints of the flux tubes where the jets are emerging. Despite the fact that $\nabla \times \textbf{B} \neq 0$ and $\frac{(\nabla \times \textbf{B})\times\textbf{B}}{\mu_{0}}\neq 0$, there should be no significant effects on the evolution of the jet, since if we calculate the curl of \textbf{B} for a two-dimensional Cartesian system, we have that $\nabla \times \textbf{B}=-\frac{\partial B_{x}}{\partial y}$ $\hat{z}$, thus
\begin{equation}
    \nabla \times \textbf{B} = -1.5 \times 10^{-13} \bar{x}B_{0}tanh(\bar{y}-3)sech^2(\bar{y}-3) \hat{\textbf{z}}. \label{rotB}
\end{equation}
For $\frac{1}{\mu_{0}}(\nabla\times \textbf{B})\times\textbf{B} =\frac{1}{\mu_{0}}\left(B_{y}\frac{\partial B_{x}}{\partial y} - B_{y}\frac{\partial B_{y}}{\partial x}\right)\hat{\textbf{x}} + \frac{1}{\mu_{0}}\left(-B_{x}\frac{\partial B_{x}}{\partial y} + B_{x}\frac{\partial B_{y}}{\partial x}\right)\hat{\textbf{y}}$, then 

\begin{eqnarray}
\frac{1}{\mu_{0}}(\nabla\times \textbf{B})\times\textbf{B} &=& (3.5\times10^{-8}B_{0}^{2}\bar{x} \textrm{sech}^{2}(\bar{y}-3)\tanh(\bar{y}-3) \nonumber \\ 
&& + 8.95\times^{-9}B_{0}^{2}\bar{x} \textrm{sech}^{2}(\bar{y}-3)\tanh^{2}(\bar{y}-3))\hat{\textbf{x}} + \nonumber \\
&& (-8.95\times^{-15}B_{0}^{2}\bar{x}^{2} \textrm{sech}^{4}(\bar{y}-3)\tanh(\bar{y}-3))\hat{\textbf{y}}. 
\label{FFC}
\end{eqnarray}

If we set the value $B_{0} = 100$ G $= 0.01$ T and $\mu_{0} = 1.256\times^{-6}$ N A$^{-2}$, in the case of Eq. (\ref{rotB}), the maximum value is around $10^{-15}$ T m$^{-1}$, while for Eq. (\ref{FFC}), the maximum value of the Lorentz force in $x$ is of the order of $10^{-12}$ N m$^{-3}$, while in the direction $y$ is of the order of $10^{-17}$ N m$^{-3}$. So the current density and the Lorentz force are negligible compared to the gravity force, balanced by the pressure of the plasma. Let us explicitly see it by calculating $\rho_{i}\textbf{g}$. Suppose we take the value of the ion density ($\rho_{i} \approx 6\times 10^{-10} \text{ kg m}^{-3}$) and the gravitational acceleration ($\textbf{g}=-274$ m s$^{-2}$ $\hat{\textbf{y}}$) at $1.3$ Mm, where we place the velocity pulse that generates one of the jets under study, we have $\rho_{i}\textbf{g}= - 1.6 \times 10^{-7}$ N m$^{-3}$ $\hat{\textbf{y}}$, i.e., is about ten orders of magnitude greater than the Lorentz force acting on $y$ direction. Therefore, due to the fact of having a much larger order of magnitude of the force of gravity ($ \rho_{i}\textbf{g} >> \frac{1}{\mu} (\nabla \times \textbf{B}) \times \textbf{B}$), we can consider that the magnetic flux tube model is very close to equilibrium, i.e., practically force-free, at $t = 0$. Then, the nature of the spicule is subject primarily to the velocity pulse and not to the currents and forces provided by the flux tube.

The magnetic field is intense at $(x, y) = (0,0)$ Mm, where the field lines open upwards, reaching a quasi-constant value after $y = 3$ Mm. We show the magnetic field lines of the flux tube in the top-right panel of Fig. \ref{mag_field_plasma_beta}. In addition, the plasma $\beta$ parameter helps estimate the ratio between ion and neutral pressures to magnetic pressure, which is defined as follows:

\begin{equation}
    \beta(x,y)=\frac{p_{i}(y)+p_{n}(y)}{B^{2}/2}.
\end{equation}

Here, the pressures $p_{i,n}(y)$ are given by equations (\ref{p_n})-(\ref{p_i}), and $B^{2}=(B_{x}^{2}+B_{y}^{2})$. We display the spatial profiles of plasma $\beta$ for both magnetic field configurations on the bottom panels of Fig. \ref{mag_field_plasma_beta}, where we observe that $\beta>1$ in the lower atmosphere (the photosphere and the chromosphere) for both cases. Otherwise, $\beta<1$ in the solar corona ($y>2.1$ Mm). Such behavior of plasma $\beta$ is consistent with a vertical dominant magnetic field, as shown in \cite{Ku_ma_2017}. 

\subsection{Perturbations}
\label{perturbations}

We perturb the hydrostatic equilibrium atmosphere, initially (at $t=0$ s), by localized Gaussian pulses in ion and neutral vertical velocities, given as

\begin{equation}
v_{y_{n,i}}(x,y, t=0) = A_{v}\exp\left(-\frac{(x-x_{0})^{2}+(y-y_{0})^{2}}{w^{2}}\right).
\end{equation}
Here, $A_{v}$ is the amplitude of the pulses, $(x_{0},y_{0})$ their positions and $w$ is their width. We locate the pulses in $x_{0}=0$ Mm and $y_{0}=1.3,1.5,1.8$ Mm for the three cases, and we hold fixed $w=0.3$ Mm and $A_{v}=100$ km s$^{-1}$, this value falls in the range of velocities of Type II spicules \citep{Pereira_2012}. 

\subsection{Numerical methods}
\label{numerical_methods}

To solve the two-fluid equations (\ref{m_i})-(\ref{B}) numerically, we employ the JOANNA code \cite{Wojcik_2019}. In all simulations, we set the Courant-Friedrichs-Levy (CFL) number equal to 0.9 and choose the third-order strong stability preserving Runge-Kutta (SSP-RK3) time integrator \citep{2010nmfd.book.....D}. Additionally, we adopt the Harten-Lax-van Leer discontinuities (HLLD) approximate Riemann solver \citep{Miyoshi&Kusano_2005} in combination with a linear reconstruction and the minmod limiter. To control numerically the growth of the solenoidal constraint condition given by equation (\ref{B}), we use the extended generalized Lagrange multiplier method \citep{Dedner_et_al_2002}. This method is robust in low plasma beta ($\sim 10^{-3}-10^{-2}$) regions, as implied in the solar corona; see, for example, the bottom panels of Fig. \ref{mag_field_plasma_beta}. 

We carry out the simulations in the domain $x\in[-5,5]$, $y\in[0,30]$, in units of Mm, in a uniform grid, which is covered by 200$\times$600 cells. Here, $y=0$ Mm represents the bottom of the photosphere. Next, we impose outflow boundary conditions at the side edges specified by $x=-5$ Mm and $x=5$ Mm. Finally, we set all the plasma quantities to their equilibrium values at the bottom and top boundaries delimited by $y=0$ Mm and $y=30$ Mm.

\section{Results of the numerical simulations}
\label{Results}

We perform a series of simulations for the case when the collisions between ions and neutrals are considered \cite{Ku_ma_2017, Braileanu}. In particular, we define the following two scenarios: 1.- Uniform magnetic field, 2.- Flux tube type configuration. For each of the two scenarios, we perform three different simulations corresponding to three different vertical locations ($y_{0}=1.3,1.5,1.8$ Mm) of the velocity pulses at the initial time, $t=0$ s. Within the range that covers the chromosphere ($0.6\leq y \leq2.5$ Mm). These velocity perturbations give rise to the jets that are of interest for this analysis (we will hereafter call them \textit{Jet$_{1}$, Jet$_{2}$} and \textit{Jet$_{3}$} for the respective values of $y_{0}$ mentioned above). In the following subsections, we describe the results of the numerical simulations under the two magnetic scenarios already mentioned.

\subsection{Uniform magnetic field}
\label{Uniform_field}

Here we implement a uniform magnetic field to observe the jets' behavior in an environment where the magnetic field lines are straight and constant. In general, the magnetic conditions of the chromosphere are complex. However, in some simulations, such as in the works \citep{Mart_nez_Sykora_2018, Gonzalez_2021}, there are a few bounded regions where some jets evolve within field lines that vary smoothly. Therefore we wanted to explore these conditions in a more general context to have as a control test a magnetically uniform environment. The latter helped to make the comparison with a more complex magnetic field that mimics a flux tube, as we will describe in subsection \ref{Flux_tube_field}.

\begin{figure}[t!]
   \centerline{\includegraphics[width=0.99\textwidth,clip=]{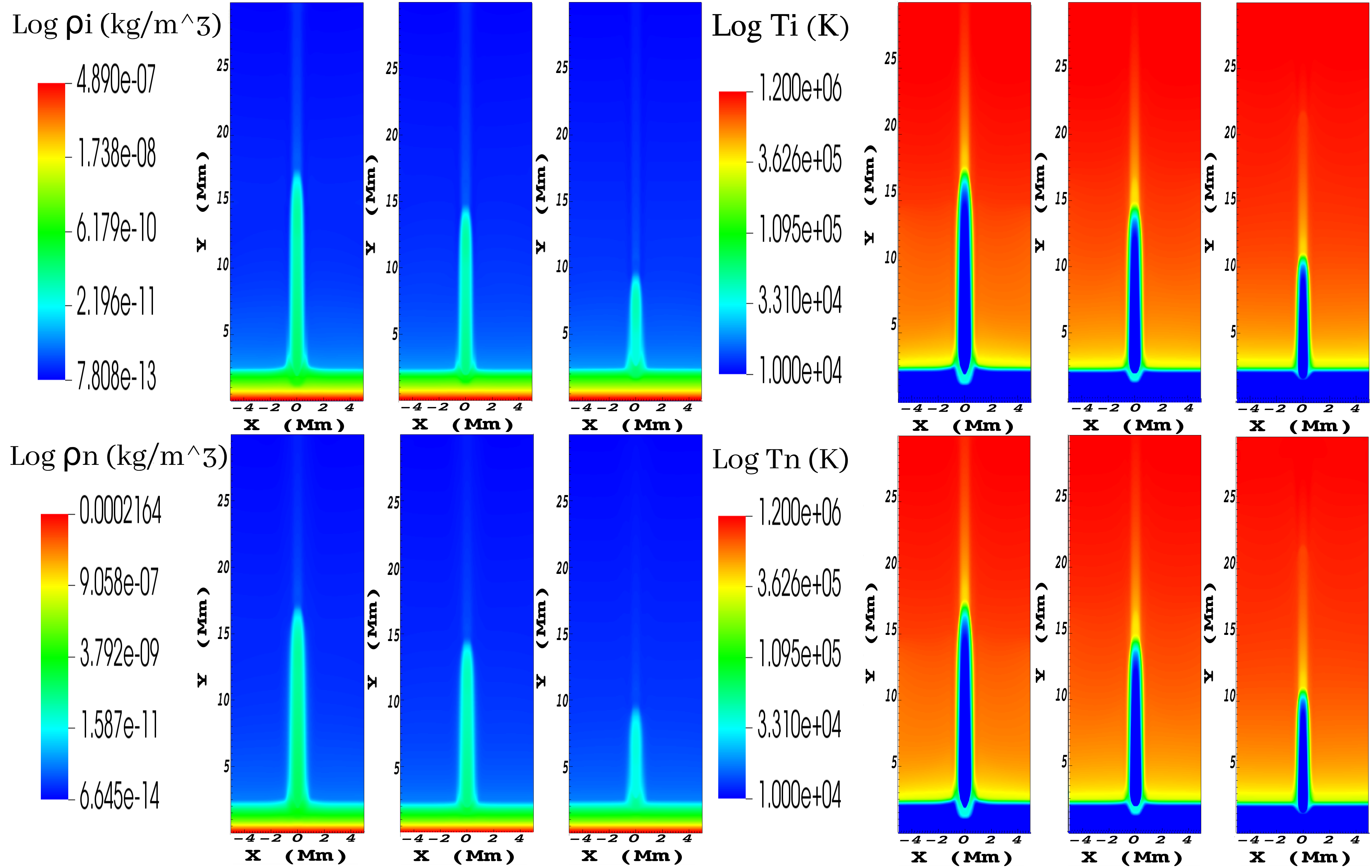}
              }
              \caption{\textbf{Uniform magnetic field case}: (Left panel) Maximum heights reached by the jets generated at different $y_{0}$. Temporal evolution of $log(\rho_{i,n}(x,y))$. Top: ion density. Bottom: density of neutrals. From left to right: jets generated at $y_{0}$=1.3 (\textit{Jet$_{1}$}), 1.5 (\textit{Jet$_{2}$}) and 1.8 Mm (\textit{Jet$_{3}$}), at t=300 s, t=270 s, and t= 210 s, respectively. (Right panel) Temperatures reached by the jets. Temporal evolution of $log(T_{i,n}(x,y))$. Top: ion temperature. Bottom: temperature of neutrals. From left to right: \textit{Jet$_{1}$}, \textit{Jet$_{2}$} and \textit{Jet$_{3}$} at t=300 s, t=270 s, and t= 210 s, respectively.}
\label{rho_const_and_T}
\end{figure}

We perform three simulations for the vertical magnetic field by launching a velocity pulse for ions and neutrals with an amplitude ($A_{v}$) of 100 km s$^{-1}$ at different vertical positions $y_{0}=1.3, 1.5, 1.8$ Mm. We set a uniform magnetic field $\textbf{B}=[0, B_{0}]$, with $B_{0}=30$ G, and we launch the pulses in a region where the condition $\beta < 1$ is satisfied (See the bottom-left panel of Fig. \ref{mag_field_plasma_beta}). The simulations were allowed to run up to a physical time of $t_{f}=600$ s. On the left side of Fig. \ref{rho_const_and_T}, we show the logarithm of the mass density for ions and neutrals, $\rho_{i,n}$ [kg m$^{-3}$]. From left to right \textit{$Jet_{1}, Jet_{2}$} and \textit{$Jet_{3}$}, respectively. At the top, we display $\rho_{i}$, while at the bottom, we show $\rho_{n}$; both quantities represent the jets. Each snapshot shows the jets' maximum heights ($h_{max}$) for both fluids. The times at which they reached their maximum heights are $t=300$ s, $t=270$ s, and $t=210$ s, respectively. For this simulation, $S_{ni} \neq 0$ allows the collision between fluids and the exchange of momentum between the ions and neutrals. Both fluids that make up the jets reach equal heights, evidencing the coupling of particles and collimating equally to the neutral part even after exceeding the times when the maximum heights are reached. As discussed in \cite{Oliver_2016}, when the charged particles act over neutrals, the species show a joint dynamicity. The temperatures in the cores of the jets remain constant throughout the evolution (see Fig. \ref{rho_const_and_T}). However, on the tips, we observe temperatures of up to 3.6 - 6.5$\times 10^{5}$ K; since ions and neutrals are coupled, they exhibit collective behavior. 

\subsection{Flux tube type configuration}
\label{Flux_tube_field}

As we have seen in some works \cite{Mart_nez_Sykora_2017, Mart_nez_Sykora_2018}, the magnetic configuration over spicule-type jets evolves as it diverges from bottom to top. For example, in \cite{Konkol_2012, Ku_ma_2017}, the authors use a general 2D expression for an open magnetic field, while in \cite{Magyar_2021}, the authors employ a general 3D expression of a flux tube. In our analysis and for this simulation stage, we use a simple analytic expression that recreates a 2D flux tube as described below. 

\begin{figure}[t!]
   \centerline{\includegraphics[width=0.98\textwidth,clip=]{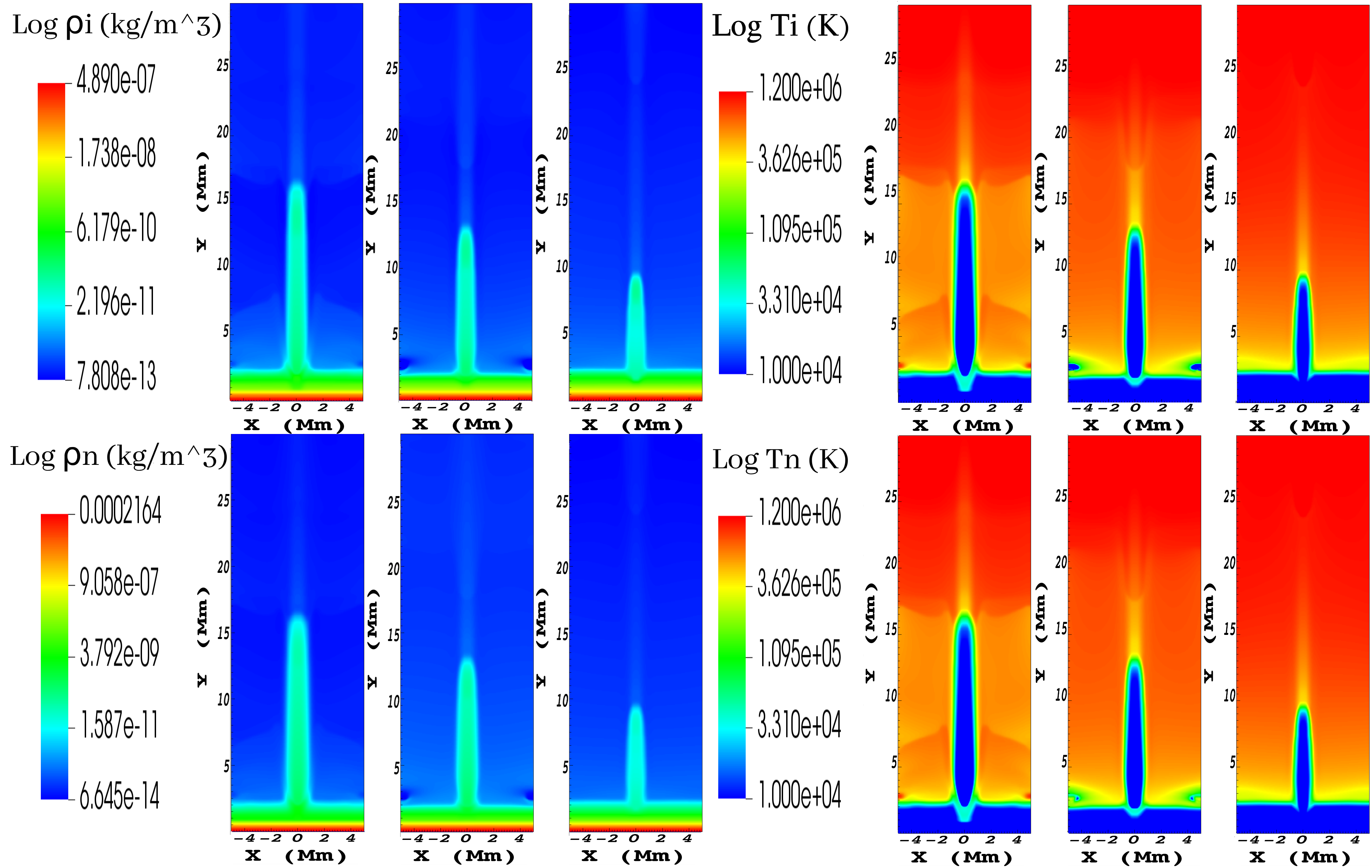}
              }
              \caption{\textit{\textbf{Flux tube case}}: (Left panel) Maximum heights reached by the jets generated at different $y_{0}$. Temporal evolution of $log(\rho_{i,n}(x,y))$. Top: ion density. Bottom: density of neutrals. From left to right: jets generated at $y_{0}$=1.3 (\textit{Jet$_{1}$}), 1.5 (\textit{Jet$_{2}$}) and 1.8 Mm (\textit{Jet$_{3}$}), at t=300 s, t=250 s, and t= 190 s, respectively. (Right panel) Temperatures reached by the jets. Temporal evolution of $log(T_{i,n}(x,y))$. Top: ion temperature. Bottom: temperature of neutrals. From left to right: \textit{Jet$_{1}$}, \textit{Jet$_{2}$} and \textit{Jet$_{3}$} at t=300 s, t=250 s, and t= 190 s, respectively.}
\label{rho_flux_tube_and_T}
\end{figure}

The jets propagate in an embedded environment in a magnetic field configuration as described in Equations (\ref{Bx})-(\ref{By}). Here $B_{0}=100$ G. This magnitude decreases with a height reaching a constant value of $0.6B_{0}$ after exceeding $y=3$ Mm. Under this flux tube, a component of the Lorentz force pulls in the $-y$ direction for the ions emerging from areas where the magnetic field is intense. Since the field becomes weaker at higher values of $y$, then the component of the acceleration in the $y$ direction is given by \cite{Jackson:100964},
\begin{equation}
  a_{y}=\frac{d v_{\parallel}}{dt}=-\frac{1}{2}\frac{v_{\perp}^{2}}{B(y)}\frac{\Delta B_{y}}{\Delta y},
\end{equation}
will become positive (here, $\frac{\partial B_{y}}{\partial y}$ is a negative term); therefore, particles leaving the lower parts of the footpoints at $100$ km s$^{-1}$, where the magnetic field is three times greater than in the zones $y > 3$ Mm, would undergo higher accelerations and therefore reach higher speeds ($\approx 197$ km s$^{-1}$) and thus, in principle, jets would reach greater heights. It is worth mentioning that the conditions under which the particles are found in this plasma do not ideally reflect the ideal behavior of a single particle as described by the theory, particularly alluding to the previous comment. This fact can be seen in Section \ref{Max_height}. In Fig. \ref{rho_flux_tube_and_T}, the \textit{Jet$_{1}$} can exceed 16.5 Mm, while \textit{Jet$_{2}$} can reach 13.5 Mm, and finally, \textit{Jet$_{3}$} reaches a maximum height of 9.75 Mm. The jets reached their maximum heights at t=300 s, t=250s, and t=190 s, respectively, following the coupling between species throughout evolution. Also, in the right panel of Fig. \ref{rho_flux_tube_and_T}, we see that the temperatures of the cores do not change in the ascent phase. However, only at the tip and peripheries of the jets is it evident the increase in temperature of around $1-3.6 \times 10^{5}$ K. We note an increase of temperature over the tips; in the two-fluid scenario, it could be produced by the velocity pulse that develops into a shock \citep[see, e.g.,][]{Gonzalez-Aviles_et_al_2022}. 

\section{Discussion}
\label{discussion}

\subsection{Maximum height of the jets}
\label{Max_height}

We have conducted a series of simulations to recreate the evolution of small-scale solar jets generated at different heights above the photosphere and under two different magnetic field conditions. We use a uniform magnetic field and a flux tube-type field. Initially, the atmosphere was hydrostatically stratified, and we perturbed with velocity pulses \cite{Murawski_2010} to generate the jets. In Figs. \ref{rho_const_and_T} and \ref{rho_flux_tube_and_T}, we show the simulations carried out only with $S_{i,n}\neq0$, however, and to identify the influence of the neutrals on the ions coupling, we also perform simulations with $S_{i,n}=0$ in Eqs. \ref{s_i}-\ref{s_n}. 

\begin{figure}[t!]
   \centerline{\includegraphics[width=1.0\textwidth,clip=]{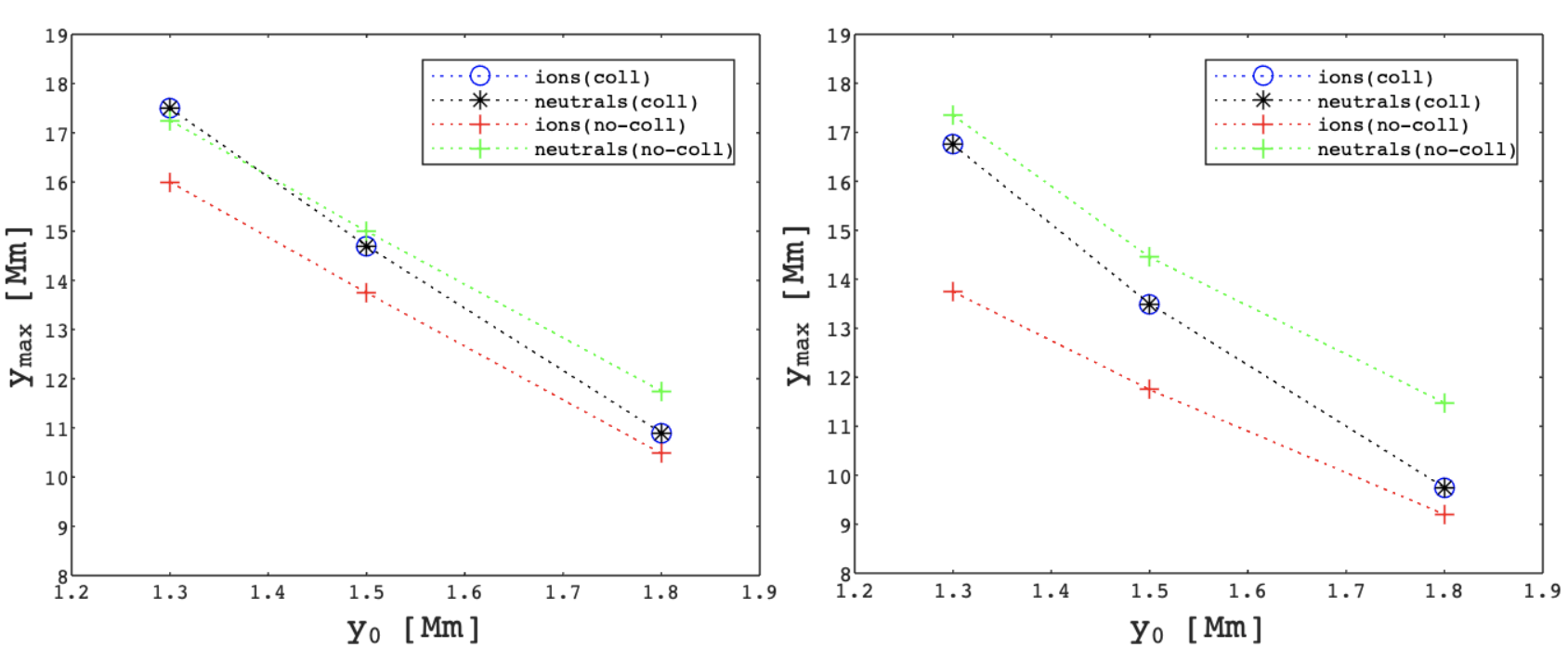}}
              \caption{\textit{\textbf{(Left) Uniform magnetic field}}: Maximum heights ($y_{max}$) of \textit{Jet$_{1}$}, \textit{Jet$_{2}$} and \textit{Jet$_{3}$} vs. vertical position of the initial velocity pulse in $y_{0}$ for $A_{v}=100$ km s$^{-1}$. Here, \textcolor{blue}{$\bigcirc$}, represents ion jets with $\textbf{S}_{in}\neq0$; $\ast$, represents neutral jets with $\textbf{S}_{in}\neq0$; \textcolor{red}{$+$}, represents ion jets with $\textbf{S}_{in}=0$ and, \textcolor{green}{$+$}, represents neutral jets with $\textbf{S}_{in}=0$. \textit{\textbf{(Right) Flux tube magnetic field}}: characters represent the same as in the previous case.}
\label{mag_const_max_h}
\end{figure}

The maximum heights reached by the ion and neutral jets whose fluids interact with each other were $y_{(i,n)max}=17.5,14.7,10.9$ Mm in the uniform field for the \textit{Jet$_{1}$, Jet$_{2}$, Jet$_{3}$}, respectively. The latter heights match those reached by macrospicules, Type I spicules, and Type II spicules for the respective jets. In the case of the flux tube $y_{(i,n)max}=16.75, 13.5, 9.75$ Mm (see Fig. \ref{mag_const_max_h}), which differed by $\Delta y_{(i,n)max}=0.75, 1.2, 1.15$ Mm; reaching greater heights the jets that evolved within the uniform field. We can also see that \textit{Jet$_{1}$}, despite having been generated in a zone ($y_{0}=1.3$ Mm) closer to the photosphere, reaches the highest maximum height compared to \textit{Jet$_{1,2}$}. The same is true for the flux tube configuration. The heights of these jets fall within the established heights for what are known as Type-I spicules (\textit{Jet$_{2}$}), Type-II spicules (\textit{Jet$_{3}$}), and macrospicules (\textit{Jet$_{1}$}). The latter happens because \textit{Jet$_{1}$} has had more mass on it than it has been able to drag as commented in \cite{Ku_ma_2017}, where they used a velocity pulse of $A_{v}=40$ km s$^{-1}$ in the case of adiabatic MHD equations. On the other hand, when the collisions between ions and neutrals turn off, the particles are no longer coupled, and the jets of different fluids present independent dynamics. The charged and neutral particle jets in the two magnetic field conditions have differences between their respective maximum heights up to $\Delta y_{(i,n)max}=1.25$ Mm for \textit{$Jets_{1,2,3}$} in the uniform magnetic field and up to $\Delta y_{(i,n)max}=2$ Mm for \textit{$Jets_{1,2,3}$} in the flux tube (see Fig. \ref{mag_const_max_h}) The latter is to outline the contribution of the neutrals on the ions through momentum transfer. 

\subsection{Temperature of the jets}
\label{Temp_jets}

At $t=0$ s (Fig. \ref{equilibrium}), the temperatures of ions and neutrals are, for $0\leq y\leq2$ Mm, $T_{i} \simeq T_{n} \simeq 10^{4}$ K. The evolution of the jets was within the range of physical time of $0\leq t_{f} \leq 600$ s. In this period, the \textit{$Jets_{1,2,3}$} reached their own $y_{max}$'s, but the orders of magnitude in temperatures remained very similar, for example, in Figs. \ref{rho_const_and_T} and \ref{rho_flux_tube_and_T}, we see that the temperature of the cores (along $x=0$ Mm) throughout their evolution did not vary significantly; this is in agreement with the observations of \cite{Beckers_1968}, \cite{Beckers_1972}, \cite{Sterling_2000}. However, on the tips of the jets, temperatures reached ($T \approx 6 \times 10^{5}$ K) above those of their bodies ($T \approx 10^{4}$ K). This result is due to the previous sweep that the shock wave produced by the pulse. At the footpoints of the jets in Figs. \ref{rho_const_and_T}-\ref{rho_flux_tube_and_T}, we see changes in temperature that appear at $t>160$ s. In the bottom of Fig. \ref{Qs_drifts} we can see that the $Q_{i}$'s are slightly greater than $0$ ($0.002 - 0.6$ W m$^{-3}$) just in these zones ($1.0\leq y \leq 3.5$ for \ref{Uniform_field} case, and $1.0\leq y \leq 1.5$ for \ref{Flux_tube_field} case). These minimum heat generation in the lower part of the jets are associated with the proportionality that exists with $(V_{n}-V_{i})^{2}$, which in turn are related to the points where the original disturbance generated. So this heat is not being produced by any event inherent in the natural evolution of jets. In the jets' lateral peripheries, we see an evident increase in temperature that keeps increasing from the beginning of the corona at $y=2.25$ Mm for $t > 0$ s, up to their $y_{max}$'s. The conversion of kinetic energy could produce this increase in temperature into heat \citep{Diaz-Figueroa_2022-1, Diaz-Figueroa_2022-2, Diaz-Figueroa_2022-3}, but it is not related to the interaction between charged and neutral particles directly, but rather to moving particles entering a relatively static environment (in ideal terms for the simulation purposes) and with temperatures exceeding $10^{6}$ K. 

In this paper, we employ a two-fluid model of simple interaction, where radiation losses and ambipolar effects of the plasma are not considered. However, it mimics the temperature behavior in the regions mentioned above of the jets. Nevertheless, there are more realistic simulations; see, for example, \cite{Mart_nez_Sykora_2017, Martinez-Sykora_2020}, where the authors analyzed the generation of spicule-jets using Cowling's conductivity employing generalized Ohm's Law, ambipolar diffusion, and multiple species with different ionization levels. 

\subsection{Collisions between ions and neutrals}
\label{Collisions_ions_neutrals}

Throughout the simulation ($t=[0-600]$ s), both in \ref{Uniform_field} and \ref{Flux_tube_field} cases, the temperature of a relatively substantial fraction of ions and neutrals inside the jets, when they have reached their $y_{max}$'s, remains constant. We calculate the characteristic collision time between fluids \cite{Oliver_2016} with $T_{i} \simeq T_{n} \simeq 10^{4}$ K, $\rho_{i}=6.3\times10^{-11}$ kg m$^{-3}$ and $\rho_{n}=1.08\times10^{-4}$ kg m$^{-3}$, using 

\begin{equation}
    \tau=\frac{1}{\nu_{ni}+\nu_{in}},
\end{equation}
where $\nu_{ni}$ is the collision frequency between ions and neutrals,

\begin{equation}
    \nu_{ni}=\frac{\alpha_{in}}{\rho_{n}},
\end{equation}

giving a value of $\nu_{ni} \simeq 384$ Hz, and since $\nu_{ni}=\nu_{in}$, then we have $\tau=1.3$ ms. As we mentioned at the beginning of this subsection, the characteristic time of the system involving the jet itself is around $6\times10^{2}$ s, so we can be sure that the collisions between particles and the exchange of momentum can keep the ions and neutrals coupled practically during the entire lifetime of the jet. The peripheries and tip of the jet, being in direct kinetic interaction with the coronal medium, reach higher temperatures of up to $T_{i} \simeq T_{n} \simeq 3.62 \times 10^{5}$ K, with $\rho_{i}=4\times10^{-12}$ kg m$^{-3}$ and $\rho_{n}=5.4\times10^{-5}$ kg m$^{-3}$. These regions' collision frequency and characteristic collision time are $\nu_{ni} \simeq 147$ Hz and $\tau=3.4$ ms. Even though the temperature is an order of magnitude higher in the outer layers of the jet, we also have an order of magnitude lower in mass densities for the center of the same jet, so the characteristic collision time turns out to be higher in the core. However, in both cases, this collision time turns out to be an order of magnitude of $10^{-3}$ s. 

\begin{figure*}[t!]
   \centering
   \includegraphics[width=14.0cm, height=12.0cm]{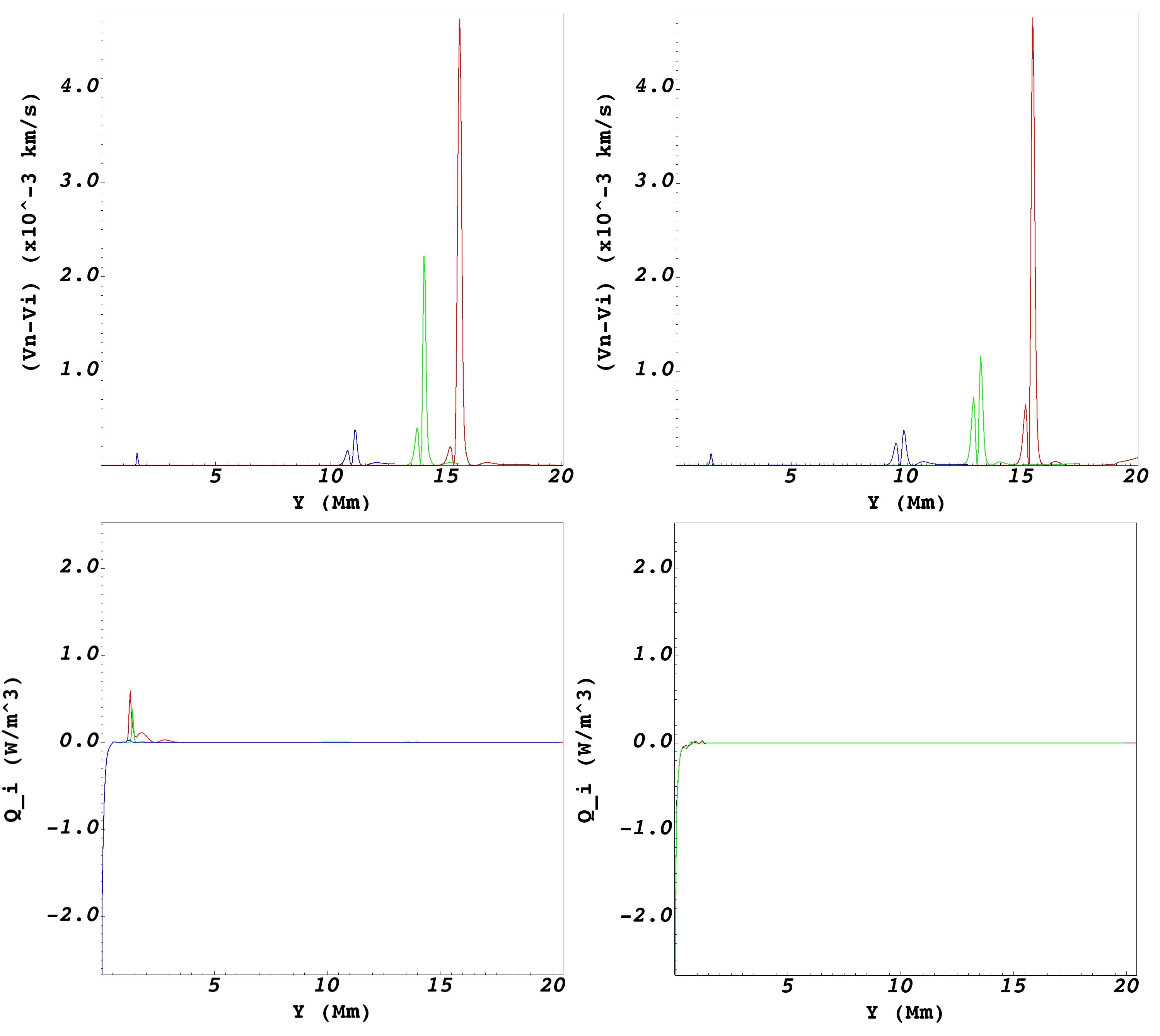}
     \caption{\textit{Top}: Difference between the neutral and ion speed (in km s$^{-1}$) evaluated along $x=0$ Mm, for \textit{$Jet_{1}$} (red), \textit{$Jet_{2}$} (green) and \textit{$Jet_{3}$} (blue) in their respective $y_{max}$'s. (Left panel: the uniform magnetic field. Right panel: flux tube.) \textit{Bottom}: Heat generation by the interaction between fluids ($Q_{i}^{i,n}$, in W m$^{-3}$) evaluated along $x=0$ Mm, for \textit{$Jet_{1}$} (red), \textit{$Jet_{2}$} (green) and \textit{$Jet_{3}$} (blue) in their respective $t_{(y_{max})}$'s. (Left panel: the uniform magnetic field. Right panel: flux tube.)}
\label{Qs_drifts}
\end{figure*}

As we can see in Fig. \ref{Qs_drifts}, the coupling of the particles is less effective at the tips of the jets than inside the jets, where the dragging of particles from the corona in relative rest ($V_{i}\approx0$ km s$^{-1}$) generates small instabilities. The order of magnitude of the velocity drifts ($V_{n}-V_{i}$) is in a range ($0-4.7\times10^{-3}$ km s$^{-1}$) for the uniform magnetic field, and ($0-4.72\times10^{-3}$ km s$^{-1}$) for the flux tube. Also, we can see a peak in the velocity drift of $0.14\times10^{-3}$ km s$^{-1}$ in $y=1.6$ Mm, near the footpoints of \textit{$Jet_{3}$}, in the uniform magnetic field, at $t=210$ s; and another peak of $0.12\times10^{-3}$ km s$^{-1}$ in $y=1.5$ Mm of \textit{$Jet_{3}$}, in the flux tube at $t=190$ s. For both magnetic field cases, the highest velocity drift is found at the tip of $ jet_{1}$.
The peaks in [$y=15.5, 14.1, 11.1$ Mm] for the \textit{$ jet_{1}, jet_{2}$} and \textit{$ jet_{3}$}, respectively, in the uniform magnetic field, are of [$0.35, 2.2, 4.7$]$\times10^{-3}$ km s$^{-1}$, the peaks in [$y=15.4, 13.6, 9.75$ Mm] for the \textit{$ jet_{1}, jet_{2}$} and \textit{$ jet_{3}$} in the flux tube are ([$0.38, 1.2, 4.72$]$\times10^{-3}$ km s$^{-1}$), not enough to have any effect on temperature.

\section{Conclusions}
\label{Conclusions_final_comments}

In this paper, we have studied the dynamics of small-scale jets in two different magnetic field configurations, using the JOANNA code to solve the Two-fluid MHD equations numerically, i.e., we consider the equations for the continuity of mass, momentum, and energy for ions and neutral particles, separately. Then, we excite the jets by launching velocity pulses at three different vertical locations $y=1.3, 1.5, 1.8$ Mm within the chromosphere range ($0.6\leq y \leq2.5$ Mm), starting from an atmosphere model in hydrostatic equilibrium. We employ a simple model that does not consider radiation losses, ambipolar diffusion, or recombination between particles. Instead, the ions and neutrals interact merely anchored to friction due to collisions. For completeness, we also study the hypothetical scenario where the collisions between the ions and neutrals are not present. The general result for this scenario indicates that momentum transfer in the jets is responsible for a relatively small increase in their maximum heights.

To summarize the results of this work, we point out the findings as follows:
\begin{itemize}
\item The \textit{$Jets_{1,2,3}$} generated within the uniform magnetic field (\ref{Uniform_field}) with $A_{v}=100$ km s$^{-1}$ showed a relationship in their maximum heights as follows, $y_{max}(Jet_{3})<y_{max}(Jet_{2})<y_{max}(Jet_{1})$, and the same was seen for the case with the flux tube (\ref{Flux_tube_field}). This behavior had already been reported in \cite{Ku_ma_2017}, where they use a velocity pulse with $A_{v}=40$ km s$^{-1}$. It is because the jets that arise from zones closer to the photosphere have a more significant amount of plasma that can be dragged by the pulse that perturbs the hydrostatic equilibrium. In this collective behavior of the plasma and under the solar conditions that we take into account, it was not possible to see any hint of what was predicted by the theory in section \ref{Flux_tube_field}. It is important to emphasize that the jets generated in the uniform magnetic field reached a higher height, with a $\Delta y_{(i,n)max}=0.75, 1.2, 1.15$ Mm, concerning their counterparts in the flux tube. This result reveals a kind of braking due to the constriction of the magnetic lines in $0\leq y \leq 5$ Mm. $Jet_{1}$ reached heights that have been reported for macrospicules \cite{Loboda_2019}, while $Jet_{2}$ and $Jet_{3}$ reached heights typical of Type I and Type II spicules \cite{De_Pontieu_2007b}, respectively. The three jets do not show similarities with surges since they are greater, less frequent, and more explosive than spicules. However, this requires a broader study to be able to determine if the $y_{0}$ at which these jets are generated can be a crucial factor in categorizing the spicules reported in \cite{De_Pontieu_2007a, De_Pontieu_2007b, De_Pontieu_2007c}, or if their nature of creation such as magnetic reconnection or another phenomenon is the one that best categorizes the jets within the family of spicules already described with current observations.
\item The characteristic times of collision between particles were $\tau = [1.3, 3.4]$ ms, for the core and peripheries of the jet, respectively, which guaranteed from $t>0$ s the coupling between ions and neutrals during the entire lifetime of the jets ($t_{f}=600$ s), so that, we observe a joint dynamic between the fluids. 
\item In Figs. \ref{rho_const_and_T} and \ref{rho_flux_tube_and_T}, we can see that the jets under the constant magnetic field are slightly thinner than those found in the flux tube configuration. The densities inside the jets remained within a value of $\rho_{i} \approx 4\times10^{-12}$ kg m$^{-3}$ and $\rho_{n}\approx 5.4\times10^{-5}$ kg m$^{-3}$) during the evolution time.
\item The velocity drifts between particles were measured at the tips of the jets when they reached their maximum heights, being negligible ($0-4.72\times10^{-3}$ km s$^{-1}$) for any heat contribution that the friction between fluids could add to the coronal region. The filamentary regions above the jets with temperatures $T_{i,n} > 6.5\times10^{5}$ K were generated by the shock wave that propagated towards the corona with velocities $V> 100$ km s$^{-1}$. The temperature of the peripheries of the jets exceeded $3.62\times10^{5}$ K, arising entirely from friction due to the collisional interaction between the coronal plasma and the jet particles.
\end{itemize}

Finally, the analysis presented in this paper complement, for example, the works of \cite{Ku_ma_2017, Oliver_2016}. In particular, this paper clearly states why the ions and neutrals behave as they are coupled. In a future study, we plan to analyze the evolution of multiple jets considering the three-fluid resistive equations, which makes it possible to be near a more realistic scenario that describe the generation of jets in the solar chromosphere.

\vspace{6pt} 



\funding{The work of JJGA is partially supported by the project CONACYT 319216 (2022), financed by "Consejo Nacional de Ciencia y Tecnología".}

\dataavailability{Not applicable.} 

\acknowledgments{We thank the anonymous referees for constructive comments and suggestions that significantly improve the clarity of the paper. The authors would like to thank the joint support from the Consejo Nacional de Ciencia y Tecnología (CONACYT), Comisión de Operación y Fomento de Actividades Académicas del IPN (COFAA), Estímulo al Desempeño de los Investigadores del IPN (EDI) and Beca de Estímulo Institucional de Formación de Investigadores del IPN (BEIFI). They would also like to thank the facilities provided by IGUM-UNAM, Campus Morelia, via JJGA., for the computer resources where the simulations were developed. The authors also want to thank the developers of the JOANNA code, which was crucial to this work. JJGA is grateful for Investigadores por México-CONACYT (CONACYT Fellow), CONACYT 319216 (2022), CONACYT LN 315829 (2021), and CONACYT-AEM 2017-01-292684 grants, which partially supported this work, along with the program "investigadores for México," project 1045 sponsor space Weather Service Mexico (SCIESMEX). We finally thank Kris Murawski for sharing the JOANNA code. Darek Wójcik developed this code with contributions from Piotr Woloszkiewicz and Luis Kadowak.}




\begin{adjustwidth}{-\extralength}{0cm}

\reftitle{References}


\bibliography{Two_fluid_jets_mdpi_2023.bib}


%


\end{adjustwidth}
\end{document}